\documentstyle[12pt,aaspp4,psfig]{article}

\def\grb{GRB\thinspace{970828}}
\def\ts{\thinspace}

\def\etal{{\it et al.}}
\def\eg{{\it e.g.,}}
\def\ie{{\it i.e.,}}

\def\lsim{\hbox{ \rlap{\raise 0.425ex\hbox{$<$}}\lower 0.65ex\hbox{$\sim$} }}
\def\gsim{\hbox{ \rlap{\raise 0.425ex\hbox{$>$}}\lower 0.65ex\hbox{$\sim$} }}

\def\arcsec{\hbox{$^{\prime\prime}$}}

\slugcomment{Submitted to The Astrophysical Journal}
\lefthead{Djorgovski \etal}
\righthead{The Afterglow and the Host Galaxy of GRB\ts{970828}}

\begin{document}

\title{\large \bf 
The Afterglow and the Host Galaxy of the Dark Burst GRB\ts{970828}
\footnotemark}

\footnotetext{Partially based on the observations obtained at the
W.~M.~Keck Observatory which is operated by the California Association
for Research in Astronomy, a scientific partnership among California
Institute of Technology, the University of California and the National
Aeronautics and Space Administration.}

\author{
S. G. Djorgovski\altaffilmark{1},
D. A. Frail\altaffilmark{2},
S. R. Kulkarni\altaffilmark{1}, 
J. S. Bloom\altaffilmark{1},
S. C. Odewahn\altaffilmark{1,3},
A.    Diercks\altaffilmark{1}
}
 
\altaffiltext{1}{Palomar Observatory, California Institute of Technology, 
                 MS 105-24, Pasadena, CA 91125}
 
\altaffiltext{2}{National Radio Astronomy Observatory, P.~O.~Box O,
                 Socorro, NM 87801}

\altaffiltext{3}{Current address: Dept.~of Physics and Astronomy,
                 Arizona State University, Tempe, AZ 85287}

\begin{abstract}

\grb\ was the first well-localized $\gamma$-ray burst (GRB) X-ray
afterglow for which no optical afterglow was found despite a prompt,
deep search down to $R_{lim} \sim 24.5$ mag.  We report the discovery
of a short-lived radio flare within the X-ray localization error
circle of this burst.  Such radio flares are seen in about 25\% of GRB
afterglows and their origin is not well understood.  The precise radio
position enabled us to identify the likely host galaxy of this burst,
and to measure its redshift, $z = 0.9578$.  The host appears to be
an interacting/merging system.  Under the assumption that the X-ray
afterglow is mainly due to synchrotron mechanism we infer the optical
afterglow flux. The observed upper limits to the optical flux are easily
explained by invoking an intervening dusty cloud within the host galaxy.
These observations support the idea that GRBs with no detectable optical
afterglows or ``dark GRBs'' can be due to dust extinction within the
host galaxies.  The census of dark GRBs can then be used to constrain
the fraction of the obscured star formation in the universe. We argue
that the existing data already indicate that the obscured
star-formation rate is no more than 1/2 of that seen at UV and optical
wavelengths.

\end{abstract}

 
\section{Introduction}
\label{sec:introduction}

Studies of $\gamma$-ray bursts (GRBs) are now one of the most active areas of
research in astronomy.  After many years of speculation based on a limited
observational evidence, the field has been revolutionized by the discovery of
long-lived and precisely located afterglows at longer wavelengths, in the
X-rays (Costa \etal\ 1997\nocite{cfh+97}), optical (\cite{vgg+97}), and radio
(Frail \etal\ 1997\nocite{fkn+97}), and the resulting determination of the
cosmological distance scale to the bursts (Metzger \etal\ 1997\nocite{mdk+97}).
 The arcminute localizations that are so essential to further progress have
been provided by the BeppoSAX satellite (\cite{pir00}) and the Rossi X-ray
Timing Explorer (RXTE) satellite (\cite{slbr+99}).  Recently the Interplanetary
Network (IPN) has started contributing a significant number of precise and
rapid localizations (\cite{hcma+2000}).  New space missions and observational
advances from the ground promise to extend this exciting track of discovery. 
For a recent review of observational results, see, \eg\ van Paradijs,
Kouveliotou \&\ Wijers (2000)\nocite{vkw00} and Kulkarni \etal\
(2000)\nocite{kbb+00}. 

Given the importance of afterglows for our growing understanding of the GRB
phenomenon, nearly every reported X-ray transient (XT) position has been
intensively followed up by ground-based facilities.  In some cases bad weather
or other such factors prevented adequate optical observations of the XT error
boxes.  However, by our accounting, in about one third of all well-localized XT
cases, no optical transient (OT) was discovered despite prompt and substantial
efforts.  
The non-detection of optical afterglows for well-localized events could be
reasonably attributed to any of the following reasons: 
($a$) a slow response (delay of many days), perhaps coupled with a steeply
      declining light curve;
($b$) a relatively shallow follow-up (\eg\ GRB 960720, GRB 970111, GRB 970616);
($c$) location in the obscured and/or crowded direction of the Galactic plane
      (\eg\ GRB 970402); or 
($d$) a relatively large error box (\eg\ GRB 970616).
 If none of these explanations for a non-detection of an optical
afterglow is viable, some physical reason must be responsible.  We
will refer to GRBs without an optical afterglow brighter than about $R
\sim 23$ mag found within at most 2 days of the event as ``dark
GRBs''.  While this definition is somewhat arbitrary, it is practical,
based on our experience and that of the community involved in the
identification and follow-up of GRB optical afterglows.

One possible explanation for such dark bursts is obscuration by dust
along the line of sight, most likely in its host galaxy
(\cite{ggv+98d}).  This would suppress the detection of an optical/UV
afterglow, but would not significantly diminish the X-ray afterglow,
and certainly not affect the detectability of a radio afterglow.  This
idea received some observational support with the discovery of a
bright radio afterglow (\cite{tfk+98}) from GRB 980329 from which
subsequently a highly reddened optical/IR afterglow was identified
(\eg\ Palazzi \etal\ 1998\nocite{ppm+98}).  If GRBs are related to the
endpoints evolution of massive stars, as envisioned in the currently
popular models, then at least some of them are expected to originate
from within dusty star-forming regions.  Thus, dark GRBs are expected
to occur, and their fraction provides an indirect, but valuable clue
to the nature of their progenitors.

\grb\ was the first clear example of a dark GRB.  The GRB was detected
and well localized by the All-Sky Monitor aboard XTE (\cite{rwsl97}).
Prompt and deep searches failed to identify the optical afterglow
(\cite{odk+97}, \cite{ggv+98d}).  Further progress in understanding of
\grb\ was stymied by the lack of an afterglow at either optical or
radio wavelengths.

In fact, we detected a weak radio flare from GRB 970828 but had failed
to appreciate its significance at the time.  A significant subsequent
development was the discovery of a radio flare from GRB 990123
(\cite{kfs+99}).  Sari \&\ Piran (1999)\nocite{sp99} attributed this
emission to the ejecta re-heated by the short-lived reverse shock.
Similar short-lived (duration of less than a few days) radio flares,
distinct from the long-lived (weeks to months) radio afterglows, have
been seen in several GRBs. 

The Japanese X-ray satellite ASCA carried out rapid observations of
\grb\ and found a prominent flare about 1.4 d after the burst
(\cite{yno+99}). The flare appeared to exhibit a feature in its
spectrum at about 5 keV. GRB 970508 also underwent a flare during
which a similar feature was identified, and Piro \etal\
(1999)\nocite{pcf+99} interpreted it as the Fe ${\rm K}_\alpha$ line
at the redshift of the host galaxy (\cite{bdkf98}). If the feature
seen in the X-ray spectrum of \grb\ is interpreted similarly, its
redshift would be $z \sim 0.33$ (\cite{yno+99}).

In this paper we present the radio observations of \grb\ and argue
that the short-lived transient in the tight ROSAT localization
(\cite{gse+97}) of \grb\ was a radio flare from \grb.  Its
sub-arcsecond localization enabled us to identify a host galaxy and
obtain its redshift, $z = 0.9578$, which forces us to reevaluate the
interpretation of the X-ray feature detected by ASCA.  The optical
observations support the suggestion that \grb\ was optically dark,
most likely due to dust extinction within the host galaxy.

The organization of this paper as follows.  The radio and optical
observations are presented in \S\ref{sec:observations}.  The
morphology and the physical properties of the host galaxy are
discussed in \S\ref{sec:host-galaxy}.  In \S\ref{sec:dust-extinction}
we investigate the physical parameters of the region causing optical
extinction along the line of sight to the GRB. The ramifications of
our findings to the understanding of the X-ray observations (flare,
spectroscopic feature) are discussed in \S\ref{sec:X-ray}.  We
conclude in \S\ref{sec:conclusions}.

\section{Observations and Results}
\label{sec:observations}

\grb\ was discovered by the All Sky Monitor (ASM) aboard RXTE on 1997
August 28.74 UT, with a duration of approximately 160 sec in the 2--12
keV energy band (\cite{rwsl97}).  Thanks to the prompt dissemination
of the initial localization, a number of missions were able to carry
out rapid follow up observations.  The afterglow position was further
constrained by ASCA observations (\cite{yno+99}) to an error circle
with a diameter of 1 arcminute, and then finally reduced by High
Resolution Imager (HRI) observations on the ROSAT satellite
(\cite{gse+97}) to a circle with 10\arcsec\ radius and centered on
$\alpha$\ =\ $18^h08^m31.7^s$ and $\delta$\ =\
$+59^\circ18^\prime50^{\prime\prime}$ (J2000).

We observed the burst position at radio wavelengths with the Very
Large Array (VLA) starting 3.5 hours after the initial $\gamma$-ray
burst. All observations were made with full Stokes parameters, using
the entire 100 MHz bandwidth which is available.  The interferometer
phase was calibrated using the nearby point source J1756+578 or
J1810+567 and the flux scale was tied to either one of the sources
J0137+331, J0542+498, or J1331+305.  A log of the VLA observations and
a summary of the results can be found in Table~\ref{tab:Table-VLA}.

On 1997 September 1 UT a radio source at a significance of 4.5-$\sigma$ level
was detected at 8.46 GHz within the 10\arcsec\ radius ROSAT error circle, as
shown in Figure 1.  The coordinates of this radio source, hereafter VLA
J180831.6$+$591851 are: 
$\alpha$\ =\ $18^h08^m31.62^s\pm{0.06^s}$, $\delta$\ =\ 
$+59^\circ18^\prime51.32^{\prime\prime} \pm{0.25}^{\prime\prime}$
(J2000; 1-$\sigma$ errors).

To verify that this source is real and not a noise fluctuation we carried out a
number of tests. We subdivided the datasets by time intervals, by hands of
circular polarization and by the two Intermediate Frequency (IF) channels.  The
source persists in each of these sub-divided datasets. So we conclude that VLA
J180831.6$+$591851 is a real source. The transient nature of VLA
J180831.6$+$591851 is readily apparent (see Table~\ref{tab:Table-VLA}), since
it is not detected prior to this time nor in subsequent observations. Possible
exceptions to this are the $\ge$2.2$\sigma$ fluctuations at 8.46 GHz on 1997
September 3 and 1997 November 3 UT. 

As discussed in \S\ref{sec:introduction} a similar transient emission
was observed from the position of GRB 990123 (\cite{kfs+99}).  By our
accounting, such radio flares (with or without the subsequent longer
lived radio afterglow) are seen in about 25\% of well studied GRBs.
Given this and the fact that the transient radio source lies within the
compact (10-arcsecond) ROSAT error circle give us confidence to suggest
that the transient radio emission is from \grb. If so, we have localized
this GRB to the sub-arcsecond position.

Our initial deep optical imaging of the RXTE error-box was done on
1997 August 29 UT, by T. Herter, O. Pevunova, F. Fang, and K. Xu, and
on 1997 August 30, 31, and September 1 UT, by C. Steidel,
K. Adelberger, and M. Kellog, at the Palomar 200-inch Hale telescope,
in excellent conditions.  Our first observation was conducted 10 hours
after the burst and subsequent three observations were conducted at
epochs 34, 58 and 82 hours.  Multiple CCD images were obtained every
night in the $R$ band, using the COSMIC (\cite{kds+98}) prime-focus
imager, with limiting magnitudes reaching $R \sim 25$ mag.  Figure 2
shows a portion of the field, with the positions of VLA and ROSAT
sources indicated.

No significant variable sources changing by more than 0.2 mag/day were
found at either the VLA position or over the entire ROSAT error circle
down to the effective limiting magnitude $R \sim 24.5$ mag in these
images (\cite{odk+97}).  Other groups, in particular Groot \etal\
(1998)\nocite{ggv+98d}, reported limits on variable objects from less
deep observations in this time interval, which are consistent with the
deeper Palomar measurements.  This placed a strong limit on any
optical transient associated with \grb, giving it one of the highest
$\gamma$-ray or X-ray to optical apparent flux ratios to date.

Preliminary astrometric solutions were obtained using positions of 12 to 15
USNO-A1.0 stars.  We found that the RT was located roughly between what
appeared as a close pair of faint galaxies (subsequently found to be a triplet
in superior Keck images; see below). 

Subsequent deep images of the field were obtained using the LRIS instrument
(\cite{occ+95}) at the W. M. Keck 10-m telescope on Mauna Kea, Hawaii, on 1997
November~1, 1998 July~18, and 1999 April~12 and 14 UT.  Multiple CCD exposures
ranging from 300 sec to 600 sec were obtained in the $R$ band, most of them in
good photometric conditions. Exposures of Landolt (1992)\nocite{lan92} standard
fields were used for magnitude zero-point calibrations.  A portion of the field
is shown in Figure 3.  The proposed optical counterpart is now clearly resolved
into three components, which we designate A, B, and C, with the brightest
component (A) corresponding to the original Palomar ID. 

An astrometric plate solution of the Keck image was performed using 22
USNO-A2.0 stars (\cite{mbc+00}). Many of the USNO astrometric stars were
saturated in our shortest Keck exposures but the centroiding is not grossly
affected.  The r.m.s.~errors of the solution are $\sigma(\alpha) = 0.11$ and
$\sigma(\delta) = 0.22$ arcsec.  The J2000 positions of the three galaxy
components are: 
$$A:~~ \alpha~ =~ 18^h 08^m 31.756^s,~~ \delta~ =~ +59^\circ 18^\prime 52.49^{\prime\prime}$$
$$B:~~ \alpha~ =~ 18^h 08^m 31.563^s,~~ \delta~ =~ +59^\circ 18^\prime 51.15^{\prime\prime}$$
$$C:~~ \alpha~ =~ 18^h 08^m 31.540^s,~~ \delta~ =~ +59^\circ 18^\prime 49.97^{\prime\prime}$$
\noindent
with an additional centering uncertainty of $\sim 0.1$ arcsec in both coordinates.

The $R$ band magnitudes of the three components measured in a 1.2
arcsec aperture are $24.4 \pm 0.2$ mag (A), $25.2 \pm 0.3$ mag (B),
and $25.1 \pm 0.3$ (C), with the errors dominated by an unknown color
term correction and aperture correction uncertainties.

To find the total uncertainty of radio source position on the optical image, we
added in quadrature the statistical uncertainties from our astrometric
solution, the statistical uncertainty of the radio position (see above), and
the 0.25 arcsec statistical uncertainty of the USNO-A2.0 to the International
Coordinate Reference Frame (ICRF; \cite{deutsch99}).  The net total uncertainty
is $\sigma(\alpha) = 0.47$ arcsec and $\sigma(\delta) = 0.34$ arcsec.  The
offset of the RT from the closest galaxy component (B) is then: $\Delta \alpha
= 0.44 \pm 0.52$ arcsec and $\Delta \delta = 0.18 \pm 0.45$ arcsec,
corresponding to a radial offset of $0.47 \pm 0.51$ arcsec.  The probability
distribution of radial offset is approximately characterized by a Rice
distribution with $\sigma_r = 0.51$ arcsec (see Bloom, Kulkarni \& Djorgovski
2001\nocite{bkd00}). 

The central galaxy component (B) now appears as the more likely host of
the RT.  However, it is intriguing that the VLA source may be located
$between$ the two optical peaks.  We are inclined to believe that the
gap separating A and B is actually a dust lane intersecting a single
galaxy, which would then naturally account for the non-detection of an
OT form this burst. However, as pointed by the referee, several GRB host
galaxies (e.g. 990506, 991216) which exhibited irregular/complicated
morphology have been resolved by HST into distinct galaxies. Thus we
cannot exclude the possibility that the GRB is located between a pair
of closely interacting galaxies.  We eagerly await {\it Hubble Space
Telescope} imaging data to resolve this morphological ambiguity.

We observed the field of GRB 970828 with NIRC (\cite{ms94}) on the
Keck I 10-m Telescope on 1998 September 6 UT.  A total of 3200 seconds
of $K_s$-band integration was obtained in $FWHM = 0.53$ arcsec seeing
and in photometric conditions.  We observed the IR standard star SJ9101
(\cite{pmk+98}) about 2.5 hours after the GRB observations in order to
find a photometric zero-point.  In the final reduced image (Fig. 3)
galaxies A and B are well detected, but galaxy C is not.  We placed
a 0.64 arcsec radius aperture about the centroid of galaxies A and B
and star 1, and applied suitable aperture corrections.  This yields
measurements of $K_s = 20.7  \pm 0.2$ mag for galaxy A, $K_s = 21.5
\pm 0.3$ mag for galaxy B, and $K_s = 18.65 \pm 0.05$ mag for star 1,
with the errors dominated by the estimated uncertainty of the aperture
correction.  The $(R - K_s) \approx 3.7$ mag color for galaxies A and
B is moderately red in comparison to a general field galaxy population
at comparable magnitudes, consistent with some intrinsic reddening.

The initial spectroscopic observations of galaxy component A were obtained
using LRIS on 1998 July 19 UT (exposure $2 \times 1800$ sec), using a
low-resolution grating with 300 lines/mm and a 1.5-arcsec wide slit, giving the
spectroscopic resolution $FWHM \approx 15$ \AA.  The slit PA was $130^\circ$,
and thus did not include components B and C. Subsequent spectroscopic
observations were obtained with LRIS on 1998 August 23 ($4 \times 1800$ sec),
1999 March 24 ($2 \times 1800$ sec), and 1999 August 13 UT ($2 \times 1800$
sec).  On these three dates, a grating with 600 lines/mm was used, giving the
spectroscopic resolution $FWHM \approx 8$ \AA, and the slit PA was $34^\circ$,
thus including all three galaxy components (A,B,C). 

In all observations, the galaxy was dithered on the spectrograph slit
by several arcseconds between the exposures, in order to improve the
sky subtraction.  Exposures of arc lamps were used for wavelength
calibrations, and the residual instrument flexure was corrected using
measurements of the night sky lines.  The typical net resulting
r.m.s. uncertainty is $\sim 0.7$ \AA\ for the UT 1998 July 19 data,
and $\sim 0.2$ \AA\ for the rest.  Most exposures were obtained with
the slit PA very close to the parallactic angle.  Exposures of
standard stars from Oke \& Gunn (1983)\nocite{og83} and Massey \etal\
(1998)\nocite{msb+88} were obtained and used for flux calibration.
All data were processed using standard techniques.

The observing conditions were mediocre for the 1998 August 23 UT data
which resulted in a very weak detection; thus, the data from this
night were not used in our analysis.  The remaining data were taken in
good to excellent conditions, and have been averaged using
statistically appropriate S/N-based weighting, after rebinning to a
common wavelength grid.

Figure 4 shows the final spectrum of the component A.  A strong emission line
is seen at $\lambda = 7297.1$ \AA, which we interpret as the [O II] 3727
doublet, and a weaker emission line at $\lambda = 7575.3$ \AA, which we
interpret as the [Ne III] 3869 line, with a weighted mean redshift $z = 0.9578
\pm 0.0001$.  The continuum shape also shows the presence of the 4000 \AA\
break and the Balmer break at this redshift. 

The component B is detected only weakly in our 600 lines/mm grating spectra,
consistent with the magnitude difference from our direct images.  However, the
[O II] 3727 line is detected at nearly the same redshift, as shown in Figure 5.
A small, but detectable velocity shift is present, corresponding to 
$\Delta v = 170$ km s$^{-1}$ in the restframe, which is consistent with either
a velocity field amplitude within a single galaxy, or a velocity difference in
a merging pair of galaxies. 

The spectrophotometric magnitudes of the component A are:  $B = 25.9$, $V =
25.05$, and $R = 24.15$ mag, uncertain by $\sim 0.2$ mag.  For the component B
we obtain $R \approx 25.5$ mag.  These numbers are in an excellent agreement
with the magnitudes measured in our direct images. 

We do not have a significant detection of the component C in our spectroscopic
data, consistent with its measured magnitude.  It can thus be an unrelated
background galaxy, but its proximity makes it more likely to be a part of the
same system as the components A and B.

\section{Physical Properties of the Host}
\label{sec:host-galaxy}

In what follows we will assume a cosmology with 
H$_\circ$=65 km s$^{-1}$ Mpc$^{-1}$, 
$\Omega_\circ=0.3$, 
and $\Lambda_\circ=0.7$.  
At $z = 0.9578$, 
the luminosity distance is $D_L = 2.08135 \times{10}^{28}$ cm, 
and 1 arcsec corresponds to 8.53 proper kpc in projection.

The $\gamma$-ray fluence as measured by the Burst and Transient experiment
(BATSE) on board the Compton Gamma Ray Observatory was 7$\times{10}^{-5}$ erg
cm$^{-2}$ (\cite{ggv+98d}). For our adopted cosmology, this implies an
isotropic $\gamma$-ray energy release E$_\gamma=1.95\times{10}^{53}$ erg. 

We assume for the foreground Galactic reddening in this direction
$E_{B-V} = 0.036$ mag (\cite{fds99}) corresponding to $A_V = 0.112$
mag and $A_R = 0.090$ mag.  All fluxes and luminosity-dependent
quantities given below have been corrected accordingly.  For the
spectroscopically determined fluxes, we give only the random errors;
we estimate an additional, systematic flux zero-point uncertainty of
$\sim 10 - 20$\%.

For the galaxy component A, the [O II] 3727 line flux is 
$f_{3727} = (1.63 \pm 0.07) \times 10^{-17}$ erg cm$^{-2}$ s$^{-1}$, 
and the observed line equivalent width is $W_\lambda = (39 \pm 2)$ \AA.  
This is fairly typical for the field galaxies in this magnitude and redshift
range (\cite{hcbp98}).  The corresponding line luminosity is 
$L_{3727} = (8.9 \pm 0.4) \times 10^{40}$ erg s$^{-1}$.  
Using the star formation rate (SFR) estimator from Kennicutt
(1998)\nocite{ken98}, we derive the 
SFR $\approx 1.2 ~M_\odot$ yr$^{-1}$, with a net uncertainty of $\sim 30$\%.

For the component B, the [O II] 3727 line flux is 
$f_{3727} = (0.45 \pm 0.1) \times 10^{-17}$ erg cm$^{-2}$ s$^{-1}$, 
and the observed line equivalent width is $W_\lambda = (40 \pm 10)$ \AA,
consistent with the magnitude difference relative to the component A.  
The estimated SFR is $\approx 0.3 ~M_\odot$ yr$^{-1}$. 

An alternative way to estimate the SFR is from the UV continuum power at
$\lambda_{rest} = 2800$\AA\ (\cite{mpd98}).  For the component A, the observed
flux at the corresponding wavelength $\lambda_{obs} \approx 5482$ \AA\ is
$f_\nu = (0.32 \pm 0.03) ~\mu$Jy, 
the restframe continuum power is then 
$P_{2800} = (8.9 \pm 0.9) \times 10^{27}$ erg s$^{-1}$ Hz$^{-1}$, 
and the corresponding SFR $\approx 1.1 ~M_\odot$ yr$^{-1}$, 
remarkably close to the estimate from the [O II] 3727 line. 

We are completely insensitive to any fully {\it obscured} components
of the SFR, so these relatively modest SFR's are only lower limits to
the true star formation rate.

The shape of a galaxy's spectrum spectrum contains some information about its
(unobscured) star formation history.  Figure 6 shows a median-binned spectrum
of the component A, along with several dust-free population synthesis model
spectra (\cite{bc93}).  The models shown illustrate two extreme cases, a
constant SFR, and a model where all of the star formation is accomplished in an
instantaneous burst.  No attempt is made here to fit any models to the data,
since there are far too many parameters which can be adjusted; the comparison
is meant to be simply illustrative.  The constant SFR models which fit the
Balmer break overproduce the UV continuum, as does a post-starburst model with
the age of 100 Myr after the burst.  However, this can probably be fixed with a
very modest amount of reddening, and the built-in no extinction assumption of
the model is unrealistic in any case.  On the other hand, post-starburst model
with larger ages does not reproduce the shape of the UV continuum just blueward
of the Balmer break.  Thus, the observed spectrum is consistent with an
actively star forming galaxy or a relatively young post-starburst galaxy with a
modest amount of reddening. 

Another interesting hint is provided by the presence of the unusually strong
[Ne III] 3869 emission line, with a total flux of about one third of the [O II]
3727 line.  This points to an unusually hot H II region, ionized by very
massive stars. The absence of other prominent emission lines allow us to
exclude the presence of a significant hidden AGN. The [Ne III] line has been
detected in some other GRB hosts (\cite{bdkf98,bdk01,dbk00}) and it provides an
indirect clue that GRBs may be connected to massive star formation.  Further
discussion of this issue will be reported elsewhere. 

From the observed continuum flux in the restframe $B$ band
($\lambda_{obs} \approx 8700$ \AA, $f_\nu = (1.23 \pm 0.11) ~\mu$Jy,
we derive the restframe absolute magnitude $M_B \approx -19.63 \pm 0.1$ mag 
for the galaxy component A.  This corresponds to a $\sim 0.5~L_*$ galaxy today.

Perhaps the most intriguing aspect of the host is the overall
morphology of the system.  It can be interpreted either as an
interacting/merging system of two or more galaxies (similar, perhaps,
to the host of GRB 980613; see \cite{dbk00}).  Alternatively, it can
be interpreted as an advanced merger, with the apparent division
between the components A and B being due to a dust lane which may be
hiding the true nucleus of the system.  In either case, a galaxy
interaction is implied, which is not uncommon for this redshift range
(see, \eg\ \cite{fal+00}).  Galaxy mergers are known to lead to bursts
of star formation, often with highly obscured components.  It is then
very intriguing that the RT position is consistent with the region
between the two brightest visible components of the overall system.

\section{The Physical Parameters of the Obscuration, and the Evidence for a Jet}
\label{sec:dust-extinction}

\grb\ was a prototypical dark GRB.  The detection of a prompt radio flare
positionally coincident with its X-ray afterglow enabled us to identify
its host galaxy and obtain its redshift.  The simplest explanation for
the non-detection of an optical afterglow is that it was suppressed due
to extinction within the host galaxy.  Here we attempt to estimate the
physical parameters of this extinction.

We first compute the expected optical flux in the framework of the
fireball model.  In particular, we assume that the 
afterglow radiation arises primarily from synchrotron mechanism.
We note, other than that for GRB 000926 Harrison \etal\ (2001)\nocite{hyb+01},
all other afterglows have been well explained within the framework
of synchrtoron emission  (\eg\ Panaitescu \&\ Kumar 2000\nocite{pk00}).
Comparing the expected value of the optical flux
to the measured upper limits
then yield an estimate of the internal extinction.  

The basic input
to the afterglow model are the X-ray afterglow observations. Following
the detection by XTE (\cite{rwsl97}), the X-ray afterglow was observed
by the PCA aboard XTE (\cite{mcc97}), ASCA (\cite{yno+99}) and ROSAT
(\cite{gse+97}). The X-ray fluxes and epochs are respectively, $10^{-11}$
erg cm$^{-2}$ s$^{-1}$, 0.16 d; $3\times 10^{-13}$ erg cm$^{-2}$ s$^{-1}$,
1.97 d; and $2.5\times 10^{-14}$ erg cm$^{-2}$ s$^{-1}$, $\sim 7$ d.
The fluxes refer to the 2--10 keV band for the PCA and ASCA but the
0.1--2.4 keV band for ROSAT. However, given the flat spectrum of the
afterglow (photon index $\sim 2$; see \cite{yno+99}) distinctions in
the band are not critical.

The decay index of the X-ray power-law decay, $f_{\nu,X} \propto t^{\alpha_X}$,
is best measured with the ASCA observations which began 1.17
d after the burst.  Yoshida \etal\ (1999)\nocite{yno+99} report
$\alpha_X=-1.44\pm 0.07$ (excluding the flare at $t\sim 1.25\times 10^5$
s; see \S\ref{sec:introduction} and below).  The decay index derived from
PCA and ASCA fluxes is $-1.4$, in excellent agreement with the $\alpha_X$
derived from ASCA data analysis. The excellent agreement of the two
decay indicies provide strong confirmation of the reality of the flare
noted by Yoshida \etal\ (1999)\nocite{yno+99}. The decay index appears
to steepen by the time ROSAT observations were conducted (epoch, 5--7 d;
Greiner \etal\ 1997\nocite{gse+97}) since $\alpha_X$ from ASCA and ROSAT
fluxes is $-1.97$. However, over the epoch of greatest interest to optical
afterglow observations, 4 hours to 2 days, the relevant $\alpha_X$ is
that measured by the PCA and ASCA.  We adopt $\alpha_X=-1.44$ for the
duration relevant to optical observations, 4 hours to 4 days.

The fireball model allows us to estimate the optical flux given the
X-ray light curve. We assume that the fireball is expanding in constant
density medium.  This assumption is justified given that all well
studied afterglows are better explained by constant ambient density
(\eg\ \cite{pk00}).  With this assumption, we can safely assume that
over the time interval of interest to us (4 hours and longer) the X-ray
band lies above the cooling break.  In this regime, the X-ray flux is
expected to undergo a power-law decay with index $\alpha_X=-(3p-2)/4$
(spherical fireball) or $\alpha\sim -p$ (a jet fireball); see Sari, Piran
\&\ Halpern (1999)\nocite{sph99}. Here, $p$ is the power law index of the
shocked electrons in the forward shock region.  We thus obtain $p\sim 2.6$
(sphere) and $p\sim 1.44$ (jet).  For all well-studied GRBs to date, $p$
appears to be tightly clustered around about 2.4 (\eg\ \cite{pk00}).
Thus we conclude that on the timescale of interest (the first 2 days)
the afterglow emission was essentially isotropic and that $p=2.6\pm 0.1$.

The expected optical spectral density at frequency $\nu_{opt}$ is $f_{\nu,opt}
= f_{\nu,X}(\nu_{opt}/\nu_X)^{-p/2}$ where $f_{\nu,X}$ is the X-ray spectral
density at frequency $\nu_X$.  From Yoshida \etal\ 
(1999)\nocite{yno+99} we find $f_{\nu,X} \sim 0.05\,\mu$Jy ($\nu_X=4\times
10^{17}\,$Hz) at $t=1.97$ d; in deriving $f_{\nu,X}$ we assumed that the
photon index of the X-ray afterglow is 2 (see above).  Thus the
expected optical spectral density is $\sim 1\,t_{\rm d}^{-1.44}\,$ mJy
where $t_{\rm d}$ is the time in days since the GRB event.  This
translates to $R\sim$ 13.5, 14.8, 16.7 and 17.5 mag, respectively at 4
hours, 10 hour, 34 hour and 58 hour after the burst. These predicted
optical magnitudes are much brighter than the upper limits of 24.5 mag
and require significant local extinction.

However, the extrapolation from the X-ray afterglow to the optical afterglow
has two major caveats.  We have assumed that the optical density follows the
X-ray spectrum and thus $f_{\nu,opt}=f_{\nu,X}(\nu_{opt}/\nu_X)^{-p/2}$ with
$f_{\nu,X} \propto \nu_X^{\alpha_X}$.  However, this assumption is true as long as
$\nu_c<\nu_{opt}$. If $\nu_{opt}< \nu_c < \nu_X$ then the optical spectral flux
density $f_{\nu,opt}=f_{\nu,X}(\nu_x/\nu_c)^{p/2} (\nu_{opt}/\nu_c)^{-(p-1)/2}$,
resulting in a decrease on the $f_{\nu,opt}$ estimates with a local minimum for
$\nu_X=\nu_c$.  Second, the optical spectral flux density can only rise to
$f_m$, the maximum flux density -- a fundamental parameter of an afterglow in
the constant-density model. 

Except for the peculiar case of GRB 970508, we know of no firm detection of the
afterglow rising to a maximum value at any wavelength other than the radio. In
all afterglows studied to date, the optical emission is found to decay in a
power law manner starting from the earliest detection.  Optical observations of
GRB 990123 were carried out within 4 hours of the burst (one of the earliest)
and even at that time the optical flux had started declining. The typical $t_m$
for the radio afterglow is days to weeks and this would imply $t_m(opt)\lsim
10^3$ s -- in agreement with the observations. If we accept that the epoch at
which the optical emission, $t_m(opt)$ is well below a day then $f_m\sim 1$
mJy. 

\noindent{\bf A Jet?}
So far, we have assumed the fireball emission is isotropic. We
justified this assumption by noting that a jet model yields too small
a value for $p$.  However, there is now a growing evidence that many
GRB afterglows have a jet-like geometry. In this respect, we note the
gradual steepening of $\alpha_X$ from $-1.44$ (epoch 4 hours to 2 days) to
$\alpha_X \sim -1.97$ (based on flux measurements between 2 d and 6 d
after the burst).  Thus the afterglow appears to be gradually
developing evidence of a jet. A fit to the X-ray light curve was made
employing the functional form in Harrison \etal\
(1999\nocite{hbf+99}), keeping $\alpha_1=-1.44$ and $\alpha_2=-2.6$,
and we derived a jet break time $t_{jet}\simeq 2.2$ days, similar to
other jet breaks (e.g.~990510; Harrison et al.~1999\nocite{hbf+99}).

The development of the jet explains why we did not see a bright radio
afterglow. In the constant-density fireball model, $f_m$ is independent of
frequency and we would expect to see the radio afterglow peak to $f_m$ on a
timescale of days to weeks. As the fireball expands laterally the afterglow
emission is weakened and the radio emission (relative to a spherical model)
suppressed. As can be seen from Table~\ref{tab:Table-VLA} and Figure 7
there is no sign of radio afterglow. The afterglow of \grb\ bears a
considerable similarity to that of GRB 990123: a strong radio flare, a weak
radio afterglow, an afterglow emission whose decay index gradually steepened on
the timescale of a few days.  It is this similarity which gives some
confidence in our simple afterglow model for GRB 970828.

\noindent{\bf Expected Optical Emission.}
We now have sufficient information to make predictions for the {\it minimum} 
expected optical flux given the X-ray detection, and the radio upper limits.  
The afterglow model in Figure \ref{fig:bbmodel} adopts a jet break time at 2.2
days, p=2.6, and $\nu_X=\nu_c$ at the time of the first X-ray measurement. 
This predicts $R\sim$ 17.5, 18.7, 20.3 and 21.2 mag, respectively at 4 hour, 10
hour, 34 hour and 58 hour after the burst.  
Thus even adopting this conservative set of assumptions, we require significant
intervening extinction, $A_{R,obs}\gsim 6.5$ mag. 
We specifically note the conservative assumption of $\nu_c=\nu_c$; if the
cooling frequency were below the optical band then the predicted optical fluxes
would be higher. In other well studied bursts, the cooling frequency is below 
the optical band at such late times.  Also, for the purpose of the discussion 
here, i.e., the absence of the optical afterglow, the geometry of the fireball
is not critical: at the epochs of interest, within the first day, spherical
fireball approximation and $\alpha_X\sim -1.44$ are excellent assumptions. 

In summary, there is no optical afterglow detected down to about $R\sim 24$ mag
(or fainter) between 4 hours and 4 nights from \grb. During this period, based
on X-ray and radio observations, we expect a minimum of $0.3\,$mJy or about 
$R \sim 17.5$ mag.  The deficit would require $A_{R,obs}\gsim 6.5$ mag.  
At the redshift of the host, central wavelength of the $R$ band corresponds
to $\lambda_{rest} \sim 3200$\,\AA.  Assuming an extinction curve similar to 
the standard Milky Way curve, we deduct the restframe extinction 
$A_{V,rest} \gsim 3.8$ mag, 
and from that the column density of $N_H(d) \gsim 6\times 10^{21}$ cm$^{-2}$ 
in the rest frame of the host galaxy; 
here we use the symbol ``$d$'' to indicate that the column density in question
was derived from dust extinction.  
This is a lower limit because our derived $A_R$ is a lower limit.
This should be compared to $2\times 10^{22}$ cm$^{-2}$, the typical column
density through Giant Molecular Clouds (GMCs) found in the Milky Way
(\cite{srb87}).  Thus a single intervening GMC could easily provide the
necessary extinction.

\section{Interpretation of the X-ray Spectrum of the Afterglow}
\label{sec:X-ray}

The X-ray afterglow of \grb\ was peculiar in two regards and here we
investigate how our measurements can help explain the X-ray observations.

\noindent{\bf X-ray Line Feature.}
Yoshida \etal\ (1999)\nocite{yno+99} found that the X-ray afterglow of 
\grb\ exhibited a strong flare at 1.25 d lasting only 2 hours.  
During this period the flux doubled up and the spectrum exhibited a feature 
at $5.04^{+0.23}_{-0.31}$ keV.
Yoshida \etal\ (1999)\nocite{yno+99} identified this with Fe K$_\alpha$
line at $z \approx 0.33$.  

Given our unambiguous measurement of the redshift, $z = 0.9578$, this
identification is no longer tenable.  Spurred by our measurement of
the redshift of the host galaxy, several authors  (\eg\ \cite{wmkr00},
Yoshida \etal\cite{yny+01}) have suggested that this feature is instead
the Fe edge (restframe energy 9.28 keV) which would appear at 4.74 keV
at a redshift of 0.9578 -- consistent with the centroid of the feature
as measured in the ASCA spectrum.  However, as has also been noted,
this identification is not
without problems since the Fe K-edge apparently is not accompanied by Fe
K$_\alpha$ line feature (at about 3.3--3.55 keV in the observer's frame).

\noindent{\bf X-ray Extinction.}
Extinction can also be measured via absorption of low energy X-rays.
At moderate to high column densities under discussion, the X-ray absorption
arises from K- and L-shell edges of oxygen and carbon.  This absorption
is essentially independent on whether the heavy elements are in gaseous
form or locked up in dust. We will refer to the column density inferred
through X-ray absorption as $N_H(X)$.  Dust extinction is due to grains whose
formation is complicated and certainly involves the refractory elements
in a substantial way.  More importantly, the size distribution of the
extinction curve is sensitive to the size distribution of grains. Since
both $N_H(d)$ and $N_H(X)$ depend on the heavy elements, it is usually
assumed that the ratio $N_H(d)/N_H(X)$ is independent of metallicity.

Thus in principle the ASCA observations should be able to independently
demonstrate the existence of a column of gas which has been invoked
to explain the lack of optical afterglow. 
As measured in the observer's frame, the expected X-ray column density is
$N_H(X;obs) = N_H(d)(1+z)^{-8/3}$ (\cite{mm83}). Thus
we expect to find  $N_H(X;obs)\gsim 1.0\times
10^{21}$ cm$^{-2}$.

Is there any evidence for an observed column density of about $10^{21}$
cm$^{-2}$  in the X-ray data? Unfortunately, the data show a complicated
behaviour.  Yoshida \etal\ (1999, 2001) noted that immediately
following the flare, the inferred column density, $N_H(X)$ shot up to
$7.1^{+3.2}_{-2.7}\times 10^{21}$ cm$^{-2}$ (here, we use the values
from the 2001 paper). The inferred column density prior to the flare
is less than $4\times 10^{21}$ cm$^{-2}$ (90\% confidence limit) which
is consistent with the optically derived value.

However, the fact that $N_H(X)$ is not constant with time means that the
post-flare  that the X-ray absorption does not occur in large
intervening structures such as clouds or GMCs and thus it is
potentially dangerous to compare $N_H(X)$ and the dust-derived
$N_H$. 

Parenthetically, we note that such variable extinction has also been
seen in the X-ray afterglow of GRB 970508 (Piro \etal\ 1999\nocite{pcf+99}).
We suggest that these large and time-variable column densities must
arise very close to the GRB i.e. within the exploding object itself
and not in the circum-burst or inter-stellar medium.

\section{Conclusions}
\label{sec:conclusions}

We have detected a short-lived radio source within the tight localization of
the X-ray afterglow of \grb.  We propose that this radio flare is similar to
the flare detected from GRB 990123 and interpreted as arising from a reverse
shock. At the position of the radio flare we identify a galaxy with a redshift
$z=0.9578$.  Its optical properties (redshift, luminosity, morphology) are
broadly similar to the other GRB host galaxies studied to date.  Its morphology
is suggestive of a merging system, and the GRB may be coincident with a large
dust lane intersecting the host. 

No optical afterglow associated with this burst was found despite prompt
searches and deep limiting magnitudes.  In the framework of the simple
afterglow theory we are able to compute the expected optical afterglow emission
using the X-ray light curve. The failure to detect optical emission is
naturally explained as due to intervening dust. An extinction of at least 7.5
mag is necessary in the observed $R$ band.  We show that a single intervening
giant molecular cloud (GMC) would provide the necessary minimum extinction. 
The GRB's possible location in a dust lane corroborates the extinction
hypothesis. 

Using the derived jet break time in \S{\ref{sec:dust-extinction}} we can
compute the opening angle of the jet as 0.12(n/1 cm$^{-3})^{1/8}$ radian (Sari
\etal\ 1999). For a two-sided jet this reduces the isotropic gamma-ray energy
of E$_\gamma=1.95\times{10}^{53}$ erg (see \S{\ref{sec:host-galaxy}}) by a
factor of 140 to 1.4$\times{10}^{51}$ erg. 

The X-ray afterglow from \grb\ was quite strong and exhibited a strong flare
during which a spectral feature was identified (\cite{yno+99}).  A similar
flare and line feature phenomenon was seen in the X-ray afterglow of GRB 970508
(\cite{pcf+99}) and the feature was identified with Fe$_{\alpha}$ at the
redshift of the host galaxy.  In the absence of any redshift information,
Yoshida \etal\ had identified the line feature in the afterglow of \grb\ with
the Fe K$_\alpha$. However, with our redshift the rest energy of this feature
is about $9.87^{+0.45}_{-0.61}$ keV -- consistent with the feature being Fe
K-edge, but the absence of an accompanying strong Fe K$_{\alpha}$ feature is
puzzling. 

The features seen in these GRBs provided considerable impetus of X-ray
spectroscopy of other afterglows.  However, we draw attention to the phenomenon
of variable $N_H$ in both these afterglows. The increased $N_H$ after the flare
is perhaps the most secure evidence to date of the presence of dense material
in the vicinity of the GRB. 

\grb\ may be the best case to date of a ``dark'' GRB, and our observations and
analysis demonstrate that it was highly extincted. 
About one third of all well-localized GRBs to date are dark, and we suggest
that most or all of them are dust-extincted GRBs.  This is not surprising,
given other evidence linking GRBs to the locations of massive stars
(\cite{bkd00}), which are predominantly found in dusty regions.  This opens a
possibility of using the GRB afterglow detection statistics to probe the
history and the relative fraction of the obscured star formation in the
universe. 

Gamma-rays can probe the densest molecular clouds and thus GRBs suffer
virtually no selection effect unlike the traditional approaches: optical/UV
(severely affected by dust) and sub-mm (sensitivity limited).  The fraction of
dark bursts ($\sim$1/3) already informs us that the obscured star-formation
rate is no more than $\sim 1/2$ of that seen at UV and optical wavelengths. 
This result is entirely independently from the constraints obtained from sub-mm
observations and diffuse cosmic backgrounds.  Further discussion of this issue
will be presented elsewhere.

The main concern in using dark bursts to probe extincted and obscured
star formation is the issue of whether GRBs or their afterglows
can significantly affect their surroundings.  Galama \&\ Wijers
(2000)\nocite{gw00} noted that a number of bursts seemingly exhibit
large column density as inferred from their X-ray afterglow (\ie\ large
$N_H(X)$) but with little or no optical extinction (as inferred from
their optical afterglow, \ie\ low $N_H(d)$), and following Waxman \&\
Draine (2000)\nocite{wd00} suggest that the GRB destroy grains along
the line of sight, thereby making it possible to see optical afterglow.
If this picture were correct, then the fraction of dark GRBs could
be higher and the dark burst fraction would not provide a stringent
constrain on the obscured star-formation fraction.

However, the variable $N_H(X)$ clearly seen in the X-ray afterglow of GRB
970828 (Yoshida et al. 2001\nocite{yny+01}) and hinted at for GRB 970508
(Piro et al. 1999\nocite{pcf+99}) cannot be reconciled with a model in
which the X-ray absorption arises out of intervening material such as
diffuse clouds or GMCs.  Indeed, the rapid variability requires that the
X-ray absorption arises very close to the GRB itself i.e. the exploding
star itself.  If so, $N_H(X)$ does not measure the true intervening
column density, unlike $N_H(d)$.  This would then explain why there
is no relation between $N_H(d)$ and $N_H(X)$; further details of this
hypothesis will be discussed elsewhere.

If our argument is correct then the fraction of dark GRBs has already
posed the strongest constraint on the total fraction of obscured star formation
in the Universe.

\acknowledgements

We are grateful to the staff of Palomar and Keck observatories for
their expert assistance during our observing runs, and to T.~Herter,
O.~Pevunova, F.~Fang, C.~Xu, C.~Steidel, K.~Adelberger, and M.~Kellog
for obtaining the Palomar images.  We thank D.~Sadava for assistance
in reduction of the infrared data. We are grateful to A. Yoshida for
discussions of the ASCA data on \grb.  We have made extensive use of
J.  Greiner's GRB web-page
(\texttt{http://www.aip.de/People/JGreiner}) and we wish to express
our gratitude to Greiner for maintaining this database.  We
acknowledge useful discussions with F. Marshall and S. Kahn.  The Very
Large Array (VLA) is operated by the National Radio Astronomy
Observatory which is a facility of the National Science Foundation
operated under a cooperative agreement by Associated Universities,
Inc.  SGD acknowledges a partial support from the Bressler
Foundation. JSB gratefully acknowledges a fellowship from the Fannie
and John Hertz Foundation.  SRK's research is supported by grants from
the NSF and NASA.  AD was supported by a Millikan fellowship at
Caltech.

\clearpage
  

\clearpage

\newpage 
\begin{deluxetable}{lrcrr}
\tabcolsep0in\footnotesize
\tablewidth{\hsize}
\tablecaption{VLA Observations of \grb\ \label{tab:Table-VLA}}
\tablehead {
\colhead {Epoch}      &
\colhead {$\Delta t$} &
\colhead {Freq.} &
\colhead {$\delta t$} &
\colhead {S$_{peak}\pm\sigma$} \\
\colhead {(UT)}      &
\colhead {(days)} &
\colhead {(GHz)} &
\colhead {(min)} &
\colhead {($\mu$Jy/ba)}
}
\startdata
1997 Aug. 28.88 &  0.1    &   4.86  & 15  &   3 $\pm$    46 \nl
1997 Aug. 31.19 &  2.5    &   4.86  & 46  &  23 $\pm$    27 \nl
1997 Aug. 31.19 &  2.5    &   8.46  & 31  &  37 $\pm$    30 \nl
1997 Sep. 01.27 &  3.5    &   4.86  & 21  &  60 $\pm$    38 \nl
1997 Sep. 01.27 &  3.5    &   8.46  & 21  & 147 $\pm$    33 \nl
1997 Sep. 01.27 &  3.5    &   1.43  & 26  &  28 $\pm$    73 \nl
1997 Sep. 03.36 &  5.6    &   4.86  & 31  &  99 $\pm$    50 \nl
1997 Sep. 03.36 &  5.6    &   8.46  & 22  & 110 $\pm$    50 \nl
1997 Sep. 16.19 &  18.5   &   8.46  & 22  &  65 $\pm$    33 \nl
1997 Sep. 19.11 &  21.4   &   8.46  & 30  &  22 $\pm$    30 \nl
1997 Sep. 23.89 &  26.4   &   4.86  & 23  & $-$18 $\pm$  40 \nl
1997 Oct. 09.13 &  41.4   &   8.46  & 137 &  20  $\pm$   14 \nl
1997 Oct. 17.06 &  49.3   &   8.46  & 118 &   7 $\pm$    18 \nl
1997 Oct. 17.73 &  50.0   &   8.46  & 76  &  20 $\pm$    21 \nl
1997 Oct. 22.24 &  54.5   &   8.46  & 80  &   7 $\pm$    21 \nl
1997 Oct. 23.00 &  55.3   &   8.46  & 44  &  19 $\pm$    50 \nl
1997 Oct. 23.24 &  55.5   &   8.46  & 54  & 40  $\pm$    36 \nl
1997 Nov. 18.99 &  82.3   &   8.46  & 127 &  32 $\pm$    13 \nl
1997 Nov. 26.99 &  90.3   &   8.46  & 142 &  27 $\pm$    14 \nl
1998 Feb. 02.73 &  158.0  &   8.46  & 100 & $-$21 $\pm$  17 \nl
1998 Mar. 05.74 &  189.0  &   1.43  & 90  & 35 $\pm$     27 \nl
1998 Mar. 24.79 &  208.1  &   1.43  & 40  & 57 $\pm$     51 \nl
1998 Mar. 29.68 &  212.9  &   1.43  & 138 &  5 $\pm$     22 \nl
1998 Apr. 27.58 &  241.8  &   1.43  &  90 & 14 $\pm$     22 \nl
\enddata
\tablecomments{The columns are (left to right), (1) UT date of the
  start of each observation, (2) Time elapsed since the GRB 970828
  event, (3) The observing frequency, (4) the total time on source,
  and (5) The peak flux density at the position of the radio transient
  on 1997 September 1, with the error given as the root mean square
  flux density.  }
\end{deluxetable}

\clearpage
\begin{figure*}
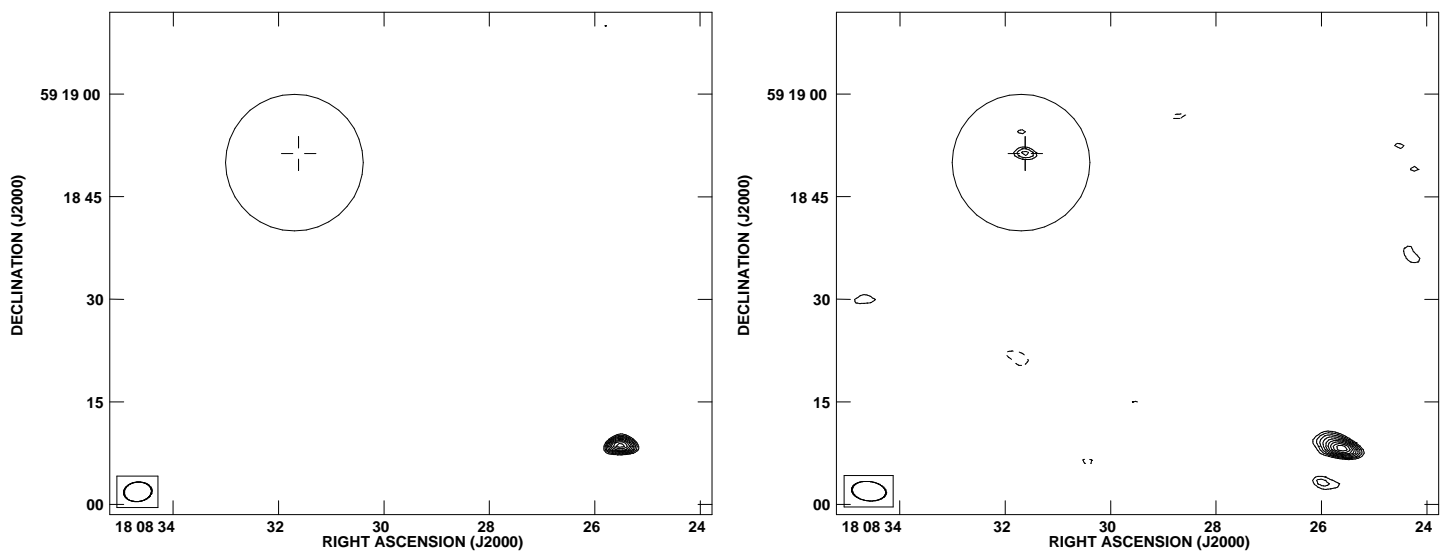

  \centerline{\hbox{\psfig{figure=g0828_aug31w.ps,angle=270,width=3.75in}
      \psfig{figure=g0828_sep01w.ps,angle=270,width=3.75in}}}
\caption[]{Radio images at 8.46 GHz toward the X-ray afterglow 
  from \grb\ on 1997 August 31 and September 1.  The 10-arcsecond
  ROSAT error radius is drawn and the ``+'' marks the location of the
  radio transient. Another radio source, seen here to the southwest,
  was originally reported by Frail \& Kulkarni (1997\nocite{fk+97b})
  since it lay inside the original ASCA and RXTE error circles.  This
  source showed no significant flux variations during our monitoring
  and is likely just a background radio galaxy. Contours are plotted
  in steps of $-3$, 3, 3.5, 4, 4.5, 5, 5.5, 6, 6.5 and 7 times the rms
  noise on 1998 September 1. The size of the synthesized beam is
  indicated in the lower left corner.\label{fig:bandaft}}
\end{figure*}

\clearpage
\begin{figure*}
\centerline{\hbox{\psfig{figure=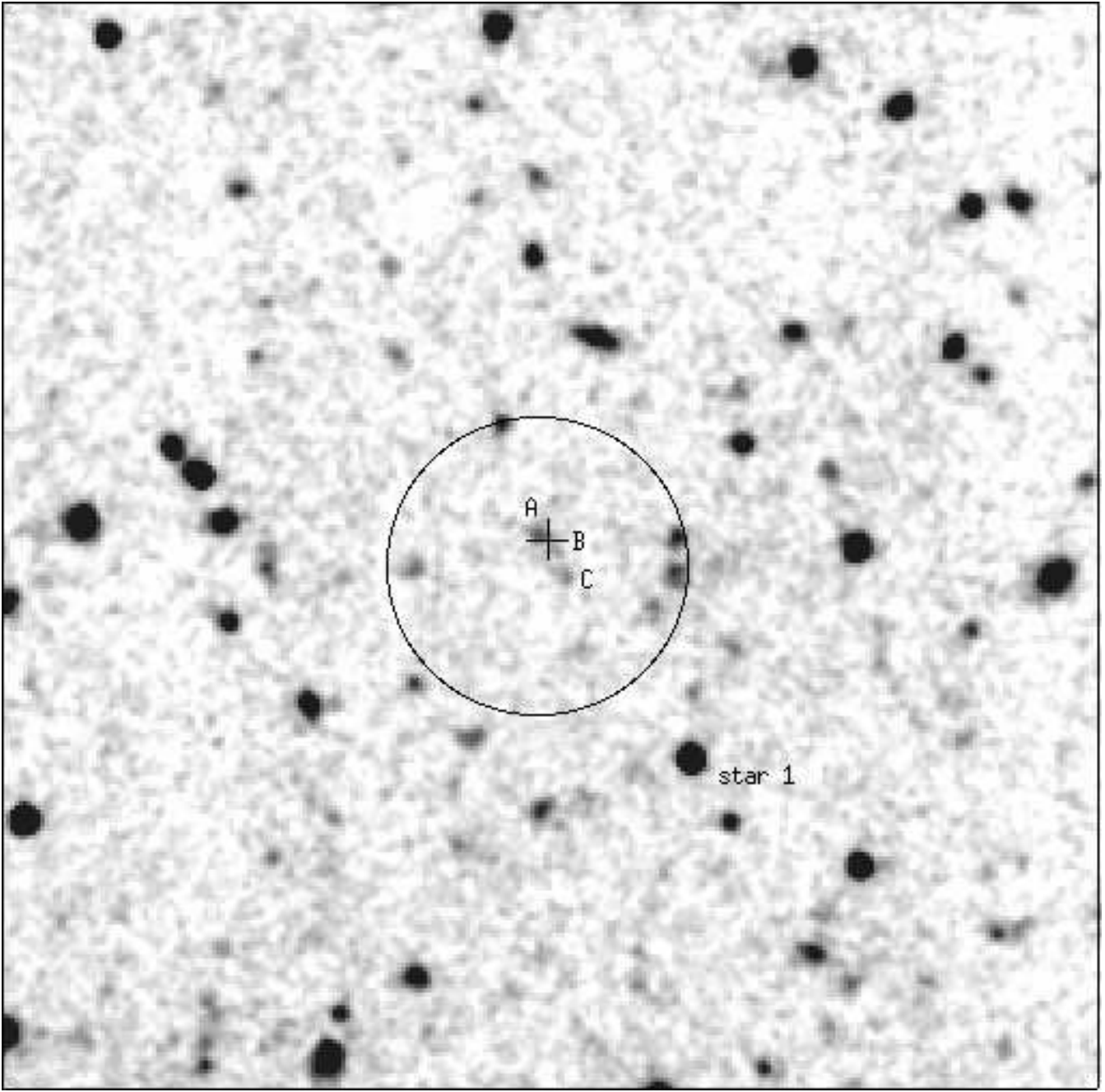,width=7.0in}}}
\caption[]{ Image of the field of \grb\ from the $R$ band data taken
at the Palomar 200-inch telescope on 1997 August 30 UT, in the $R$
band.  The field size shown is 72.5 arcsec square, with North up and
East to the left.  The ROSAT error circle of the X-ray afterglow, with
a 10\arcsec\ radius is shown.  The position of the radio afterglow is
indicated by the cross.  Proposed host galaxy components (A, B, C) are
indicated.  The offsets from star 1 to the brighter component of the
host galaxy component A are: 10.1\arcsec\ E, and 14.9\arcsec\ N.}
\label{fig:fchart1}
\end{figure*}

\clearpage
\begin{figure*}
   \centerline{\hbox{\psfig{figure=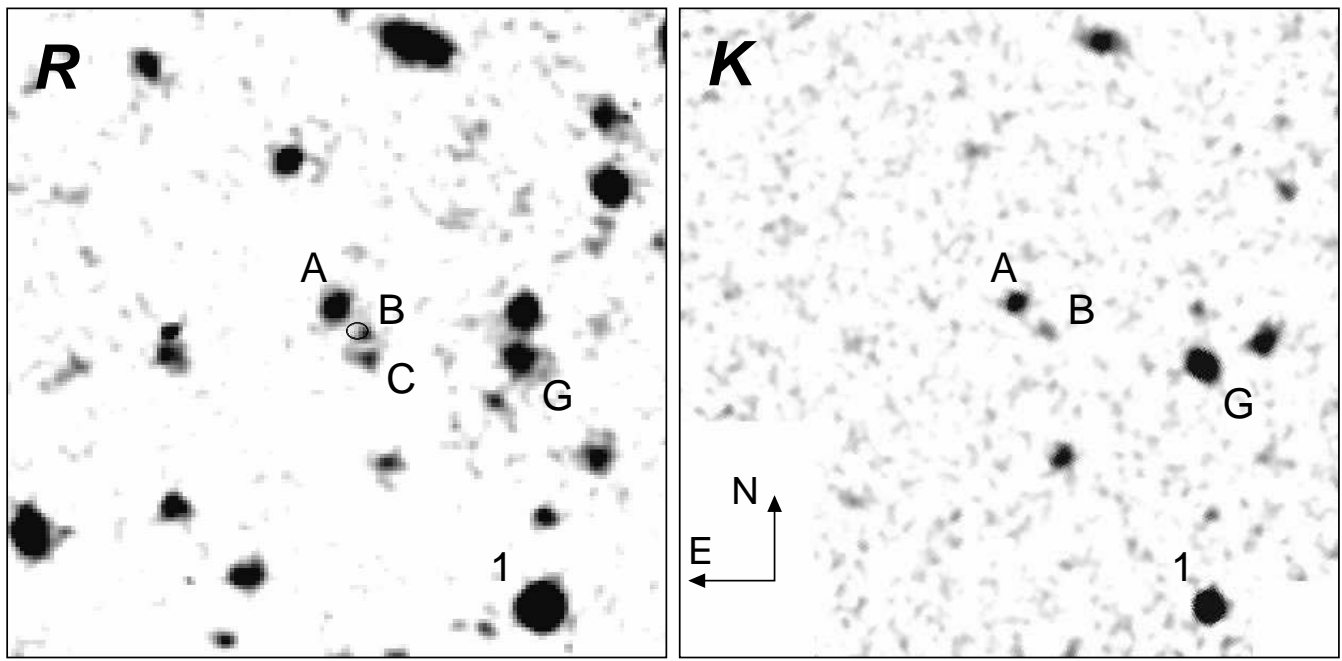,width=7.0in,angle=270}}}
\caption[]{
Close-up region of GRB 970828 in the $R$ band (left) and $K_s$ band
(right).  The images are 33 $\times$ 33 arcsec$^2$ with North up and East
to the left.  Galaxies A, B, C, and G (see text) are labeled as well
as the offset star 1.  The small ellipse at the center of the $R$ band
image is the 1-$\sigma$ error contour of the position of the radio
transient.  The transient appears nearly coincident with galaxy B but may
also have arisen in the region between galaxies A and B, potentially a
dust lane intersecting a single, larger galaxy or a merging system of
with components A, B, and possibly C.
A comparison of the $R$ and $K$ images demonstrates the red colors of
galaxies A and B; in contrast, galaxy C appears to be very blue.
Galaxy G is the very red object noted by \cite{kes97}.
\label{fig:fchart2}}
\end{figure*}

\clearpage
\begin{figure*}
  \centerline{\hbox{\psfig{figure=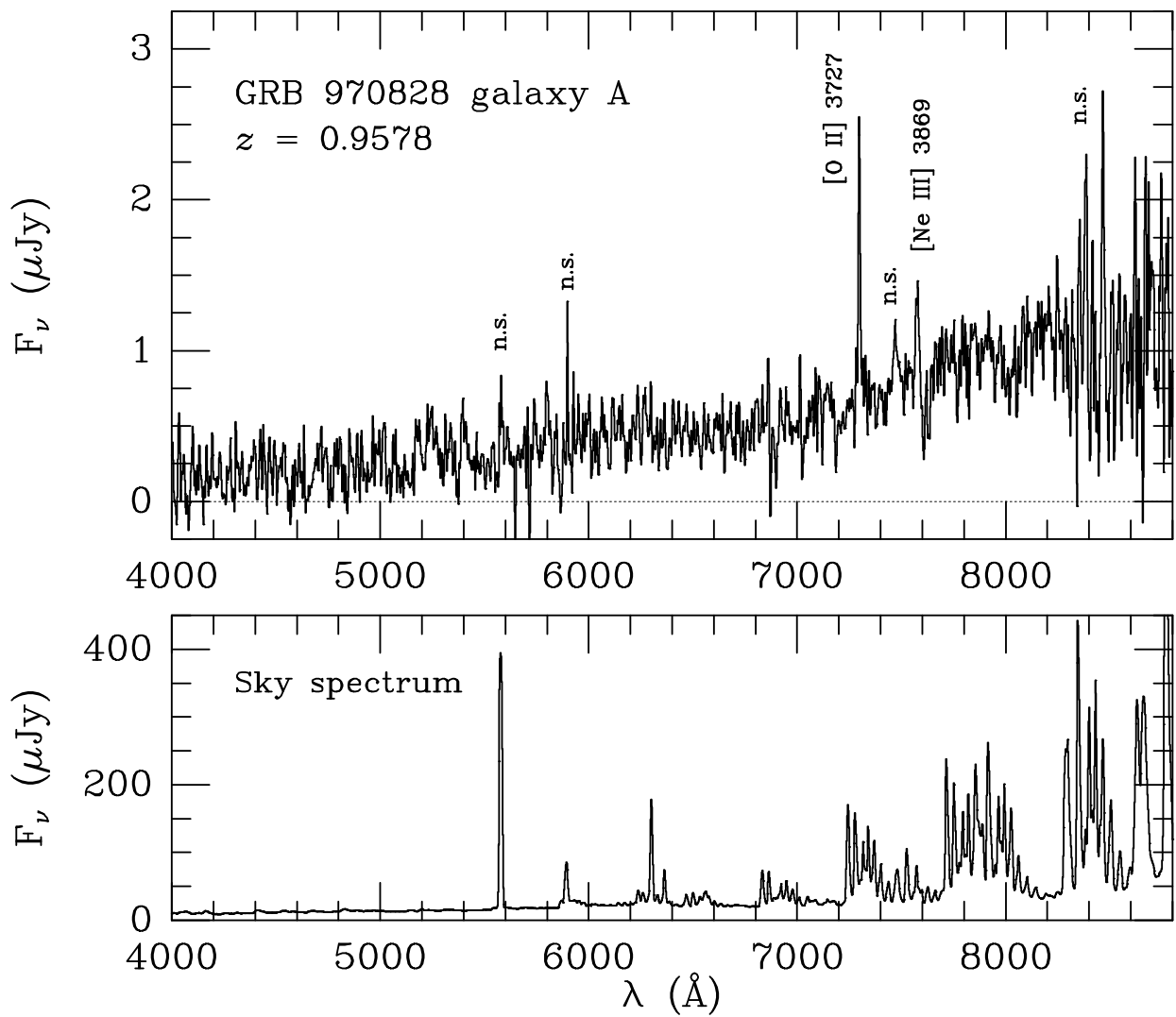,width=7.0in}}}
\caption[]{
A weighted average spectrum of the component A of the host galaxy of \grb\
obtained at the Keck.  Prominent emission lines of [O II] 3727 and [Ne III]
3869 are indicated; the remaining apparent features are artifacts of night
sky subtraction, with the sky spectrum shown in the bottom panel for
comparison.  In addition to the emission lines, the spectrum also shows
the Balmer break at the same redshift, $z = 0.9578$.
\label{fig:ospectrum}}
\end{figure*}

\clearpage
\begin{figure*}
  \centerline{\hbox{\psfig{figure=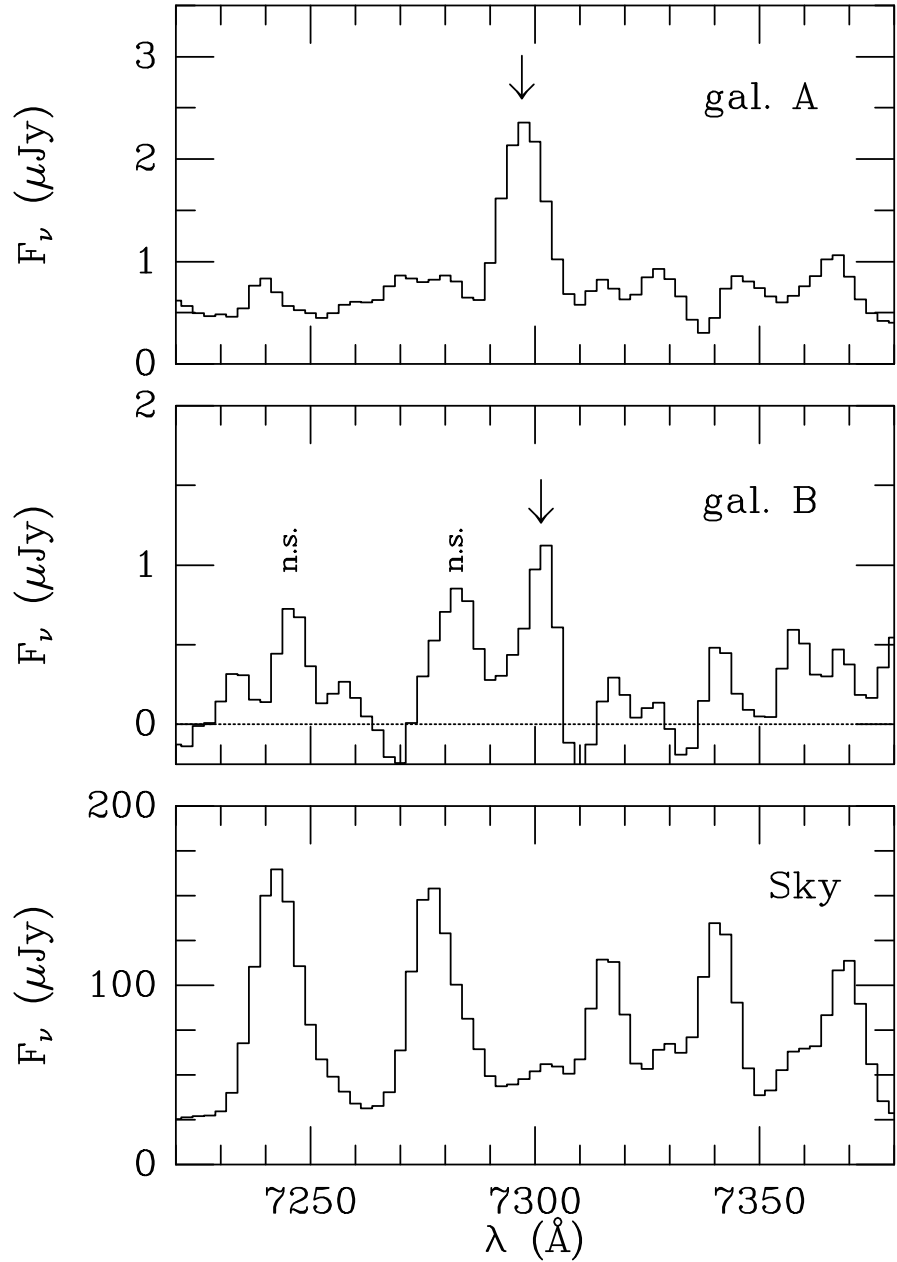,width=5.0in}}}
\caption[]{
Top: a zoom-in on the spectrum of galaxy A, centered on the [O II] 3727 line,
indicated with the arrow.  
Middle: the same, but for galaxy B, showing a weak, but definitely real line
at approximately the same wavelength.  The line falls between strong night
sky lines, and is thus not affected by poor sky subtraction.
Bottom: the corresponding portion of the night sky spectrum, shown for
comparison.
\label{fig:compspectrum}}
\end{figure*}

\clearpage
\begin{figure*}
  \centerline{\hbox{\psfig{figure=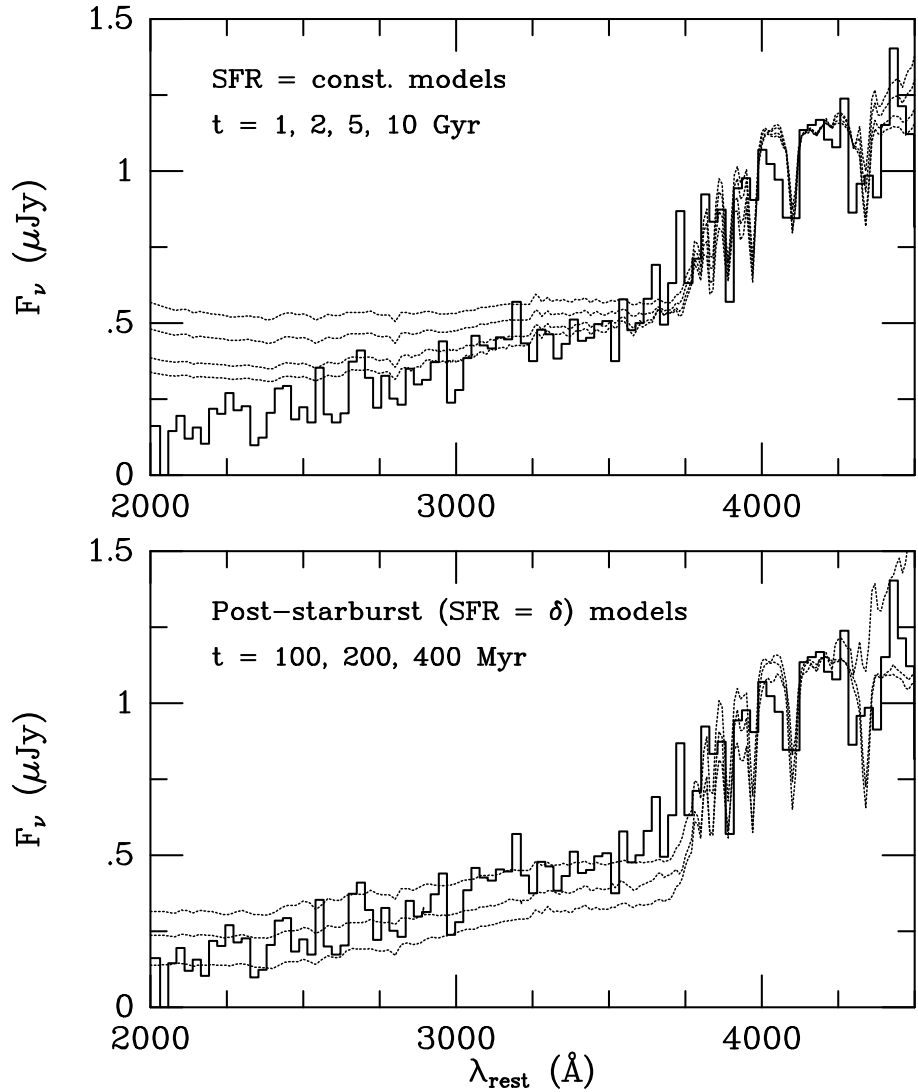,width=5.0in}}}
\caption[]{Median-binned portion of the host galaxy spectrum (solid histogram)
near the Balmer decrement, plotted in the galaxy's restframe.  Top panel shows 
Bruzual \& Charlot (1993)\nocite{bc93} galaxy synthesis models (dotted lines)
with a constant star formation, at ages 1, 2, 5, and 10 Gyr, and no extinction.  
Bottom panel shows synthesis models assuming an instantaneous ($\delta$
function) burst of star formation after 100, 200, and 400 Myr.  These
post-starburst models provide a marginally better fit to the data, but allowing
for some extinction could make the constant SFR models fit as well. 
\label{fig:modspectrum}}
\end{figure*}

\clearpage
\begin{figure*}
  \centerline{\hbox{\psfig{figure=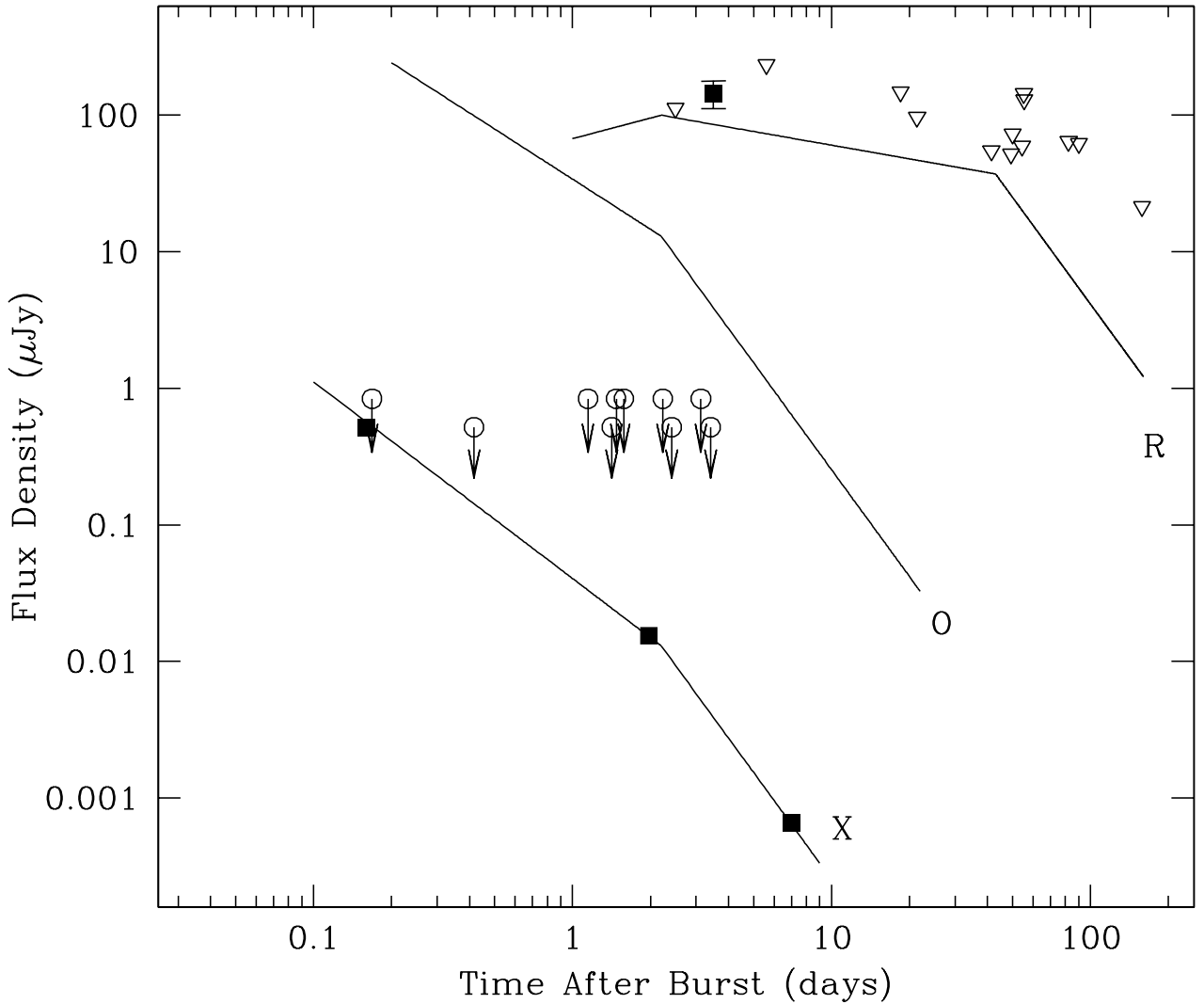,width=7.0in}}}
\caption[]{Broad-band light curves of \grb. Detections are plotted as
  solid squares. Radio (8.46 GHz) non-detections are indicated as open
  triangles and are plotted as the peak flux density at the location
  of the afterglow plus two times the rms noise in the image. Optical
  upper limits (R band) are taken from this paper and Groot \etal\ 
  (1998)\nocite{ggv+98d}, and are indicated by open circles. The solid
  lines are model light curves with a jet break $t_{jet}\simeq 2.2$
  days, as determined from a fit to the X-ray data.  The optical light
  curve is a (conservative) prediction, constrained by the X-ray and
  radio measurements and assuming that the cooling frequency $\nu_c$
  is near the X-ray band (see text for more
  details).\label{fig:bbmodel}}
\end{figure*} 

\end{document}